\documentclass[preprint]{aastex}

\def\degree{\ifmmode {^\circ}\else {$^\circ$}\fi}
\def\rstar{\ifmmode {\, R_{\star}}\else $R_{\star}$\fi}
\def\msol{\ifmmode {\, M_{\odot}}\else $M_{\odot}$\fi}
\def\rsol{\ifmmode {\, R_{\odot}}\else $R_{\odot}$\fi}
\def\lsol{\ifmmode {\, L_{\odot}}\else $L_{\odot}$\fi}
\def\msolyr{\ifmmode {\,M_{\odot}\,{\rm yr}^{-1}}\else $M_{\odot}\,{\rm yr}^{-
1}$\fi}
\def\mdot{\ifmmode {\,\dot{M}}\else $\dot{M}$\fi}

\def\mdotyr{\ifmmode {\,\dot{M}\,yr^{-1}}\else $\dot{M}\,yr^{-1}$\fi}

\begin{document}

\title{Interstellar Polarization in M31$^1$}

\author{Geoffrey C. Clayton$^2$, Michael J. Wolff$^3$, Karl D. Gordon$^4$,
Paul S. Smith\altaffilmark{4}, 
Kenneth H. Nordsieck$^5$, and Brian L. Babler$^5$} 

\altaffiltext{1}{ A portion of the results presented here made use 
of the Multiple Mirror
Telescope Observatory, a facility operated jointly by the University of Arizona
and the Smithsonian Institution.}

\altaffiltext{2}{Department of Physics \& Astronomy, Louisiana State 
University
   Baton Rouge, LA 70803; email: gclayton@fenway.phys.lsu.edu}

\altaffiltext{3}{Space Science Institute, 1540 30th Street, Suite 23,
Boulder, CO 80303-1012; email: wolff@colorado.edu}
\altaffiltext{4}{ Steward Observatory, University of Arizona, Tucson, AZ 
85721;
E-mail: kgordon, psmith@as.arizona.edu}

\altaffiltext{5}{Space Astronomy laboratory, University of Wisconsin, Madison, WI 53706; 
email: khn, brian@sal.wisc.edu}

\begin{abstract}
The wavelength dependence of interstellar polarization due 
to dust in M31 has been observed along four sightlines. 
Only one sightline had been measured previously.
The globular clusters, S78,
S150, S233 and Baade 327 were used as point sources to probe the 
interstellar dust in M31.
The Serkowski law produces good fits for all the sightlines
although the relationship between K and $\lambda_{max}$ may be different from that found 
in the 
Galaxy.
The results of this study imply that the slope K/$\lambda_{max}$ may be significantly larger
in M31.
The Serkowski curves are significantly narrower than those of the 
same $\lambda_{max}$\ in the Galaxy and may require
extreme modifications to
the size distributions of silicate particles.
The fits for the four sightlines 
reveal values of $\lambda_{Max}$ ranging from 4800 to 5500 \AA.
These are consistent with average
values of $\lambda_{max}$ measured in the 
Galaxy and the Magellanic Clouds.
The range measured for M31 corresponds to R$_V$ values of 2.7 to 3.1.
The range in R$_V$ seen in the Galaxy is 2.5 to 5.5 implying,
for this small sample, that the average size of
interstellar grains in M31 is typically smaller than that 
seen for Galactic grains if the nature of 
the grains is the same.
Also, the polarization efficiency for these sightlines is large although
some bias is expected since sightlines known to have significant interstellar 
polarization were selected
for the sample.

\end{abstract}


\section{Introduction}
There is an average Milky Way extinction relation,  depending on only one parameter, the 
ratio of total-to-selective extinction, R$_V$, which is applicable to a wide range of 
interstellar dust environments (Cardelli, Clayton, \& Mathis 1989 (CCM)). Similarly, there is an 
interstellar polarization relation, correlated with a single
parameter, $\lambda_{max}$,
the wavelength at which the interstellar polarization is a maximum (Serkowski, Mathewson,
\& Ford 1975; Clayton et al. 1995; 
Martin, Clayton, \& Wolff 1999).  The parameters, R$_V$ and $\lambda_{max}$,
both involve integrals over the size distribution of interstellar dust grains
and are well correlated with each other (Serkowski et al. 1975; 
Whittet \& van Breda 1978; Clayton \& Mathis 1988).
The existence of these 
relations, valid over a large wavelength interval, suggests that the environmental processes 
that modify the grains are efficient and affect all grains. However, the CCM relation is 
not a universal law.  It does not fit the observed extinction in the sightlines observed 
in the Magellanic Clouds (e.g., Clayton et al. 1996;
Gordon  
\& Clayton 1998; Misselt, Clayton, \& Gordon 1999; Gordon et al. 2003).  
The differences in observed extinction properties 
seen among the Local Group galaxies may be due to various environmental factors including 
metallicity.
The metallicity of the LMC and SMC is lower than the Galaxy.
Therefore, the nature of the dust in M31 
is of great 
interest because its metallicity is similar to that in the Milky Way.  

The nature of interstellar dust in M31
is summarized in Lequeux (2000). 
The UV extinction data for M31, 
which
are based on only a few lightly reddened sightlines, 
show a CCM-like far-UV extinction and a weak 2175 A bump 
(Bianchi et al. 1996). 
Several studies have suggested that the extinction in M31 in the U-band differs from
the Galaxy (Searle 1983; Iye \& Richter 1985; Massey et al. 1995). 
It has also been suggested that there is a systematic change in the U-band extinction
with distance from the center of M31 (Iye \& Richter 1985; Freedman \& Madore 1990).
Various extimates of R$_V$ ranging from 2.5 to 3.8 have been made using individual stars
and clusters in M31 (Kron \& Mayall 1960; Sharov \& Lyutyj 1989; Nedialkov \& Veltchev 1999).
Recently, Barmby et al. (2000) find 
that the U- to K-band extinction curve in M31, derived from a large number of
globular clusters, is 
consistent with CCM with R$_V$=3.1.

The interstellar polarization in the LMC and SMC shows a
similar 
wavelength dependence to that seen in the Milky Way and can be well fit by the 
Serkowski relation (Clayton, 
Martin, \& Thompson 1983; Clayton 
et al. 1996; 
Rodrigues et al. 1997). Some interstellar polarization measurements in M31 have been made 
using 
globular clusters as
point 
sources (Hiltner 1958; Martin \& Shawl 1979, 1982).
However,the wavelength dependence of interstellar polarization has been 
measured along only one
line of 
sight (Martin \& Shawl 1979).  This polarization measurement is consistent with the Serkowski 
relationship and a 
small value of $\lambda_{max}$.  
We report further observations of the wavelength dependence of interstellar polarization in
M31.  

\section{Observations}
Optical spectropolarimetry was obtained 
using the 6.5 m telescope of the
Multiple Mirror Telescope (MMT) Observatory located on Mount Hopkins, Arizona. 
Observations were made of 
four globular clusters listed in Table 1
on 2003 October 30 UT. 
All of the observations made use of the CCD Spectropolarimeter (Schmidt, Stockman, \& 
Smith 1992), upgraded with a 1200 x 800 pixel, thinned, antireflection-coated, 
UV-sensitized CCD, and an improved camera lens and half-wave plate. 
All observations utilized a 600 line mm$^{-1}$ grating for high throughput and wide 
spectral coverage, 4100-8300 \AA. The FWHM resolution with this grating 
and a slit width of 1\arcsec~is $\sim$17 \AA\ (3 pixels). 
A polarimetric measurement sequence involves four separate exposures 
that sample 16 orientations of the semiachromatic half-wave plate and totalled 
480-1440 s of integration for each of the four clusters 
through light cloud cover. 
Polarization position angles were 
referenced to the equatorial system using interstellar 
polarization standards (Schmidt, Elston \& Lupie 1992). The spectral 
flux distribution is also obtained in the course of the observations and is calibrated 
relative to spectrophotometric standard stars selected from the IRAF database 
(Massey et al. 1988).
The data have been binned and are
plotted in Figure 1 along with the spectra of the globular clusters.

In addition, spectropolarimetric observations of S78 
were obtained with the HPOL spectropolarimeter 
on the 3.5 m Wisconsin-Indiana-Yale-NOAO
(WIYN) telescope on Kitt Peak. The spectropolarimeter is a modified Boller \&
Chivens spectrograph with a rotating superachromatic waveplate after the slit 
(Nordsieck \& Harris 1996; Wolff, Nordsieck, \& Nook 1996). Spectra are obtained at eight waveplate
angles on a Reticon 400 x 1200 CCD. Two gratings are used, producing spectra and 
3 pixel resolutions of 3200-6200 \AA~(9 \AA) and 6000-10500 \AA~
(12 \AA), respectively. Two polarized star spectra and two polarized sky spectra 
are obtained 
simultaneously. The calibration accuracy in polarization is $\pm$p/500 and
in position angle is $\pm$1\degree. There is an uncertainty of 0.1\% in the 
correction for 
instrumental polarization. This small uncertainty does not affect the
results presented here.
The WIYN spectropolarimetric data were obtained in 1996 December and 1998 October. 
The four nights of data on S78 have been combined and binned
to increase the signal-to-noise.
They are plotted in Figure 1. 

The polarization measured toward M31 must be corrected for a foreground component.
These lines of sight pass through the disk and halo
of the Galaxy, and then the halo and disk of M31. However, since M31 lies
at a Galactic latitude of -22\degree, the foreground
extinction due to dust in the Galaxy is small. Several estimates have been made of the 
reddening foreground to M31. They all indicate that E(B-V) $\sim$ 0.08-0.11 
(Bianchi et al. 
1996 and references therein). Using the well-known Galactic relationship between 
reddening and interstellar polarization, the maximum foreground polarization 
is $\sim$ 0.9\% (Serkowski et al. 1975). Martin \& Shawl (1979)
estimated the actual foreground polarization by measuring a sample of Galactic stars lying in the 
direction 
of M31 as well as the nucleus of M31 itself. The foreground dust is patchy so there is 
no way to
know a priori what the foreground component is along any given sightline. 
So, Martin \& Shawl (1982) 
adopted a best-guess value of p$_{max}$=0.4\%, 
$\lambda_{max}$= 5500 \AA ~with a 
position angle of 90\degree~by averaging several foreground sightlines. 

The new 
observations reported here have been corrected for foreground polarization
in the same manner as Martin \& Shawl (1982).
A Galactic Serkowski curve, as described above, 
was assumed and the foreground polarization at 
each wavelength was calculated and subtracted.  
The foreground-corrected polarizations are plotted in Figure 1.
The observed and foreground corrected values are listed in Tables 2-6. 
The new data are consistent with the Martin \& Shawl (1982) data for S78 
in the overlapping wavelength
range. Similarly, the three wavelength
bins for S150 and one bin for S233, agree with our new data within 2$\sigma$. There is only 
one filterless observation of Baade 327 having low S/N. 
The Martin \& Shawl filter data are plotted for comparison in Figure 1. 

\section{Discussion}

The interstellar polarization due to dust in M31 
was first measured by Hiltner (1958) using 
globular clusters as
point 
sources. He made unfiltered observations of 
about 20 clusters finding several with significant amounts of polarization. 
Filter
polarimetry of a similar number of clusters was obtained
by Martin \& Shawl (1979,1982). 
Only one sightline in their sample, toward S78, had enough wavelength coverage for a
Serkowski fit to be made. 
That sightline showed p$_{max}$ = 1.98$\pm$0.07 \%, and $\lambda_{max}$ = 
3500$\pm$300 \AA\
using the Wilking et al. (1980) formulation of the Serkowski relation.

Our new data for S78 and three additional 
clusters cover approximately the same wavelength
range (4100-8300 \AA) as Martin \& Shawl but with many more data bins and smaller 
uncertainties. 
We performed nonlinear least-squares fits to the foreground corrected data for all 
four clusters using 
the Serkowski law (Serkowski et al. 1975; Wilking et al. 1980; Wilking, 
Lebofsky, \& Rieke 
1982; Whittet et al. 1992). 
We found that the  K= 1.66$\lambda_{Max}$ relationship, that holds for Galactic interstellar polarization,
does not produce good fits for the M31 sightlines (Whittet et al. 1992).
However, the Serkowski law with K as a free parameter does produce 
good fits for all of the sightlines. 
The three fit
parameters are listed in Table 1. 
The relationship between K and $\lambda_{max}$ is shown in Figure 2.
Although, we do not have enough sightlines to define the K-$\lambda_{max}$ relationship
for M31, Figure 2 shows that the slope may be steeper than in the Galaxy.

The K-$\lambda_{max}$ relationship for the M31 globulars is very
interesting in that it produces distinctly narrower ``Serkowski curves''
than those of the same $\lambda_{max}$\ in the Galaxy.  As has been shown
many times in the past 
(Wolff et al. 1994; Clayton et al. 1995; Martin et al. 1999), 
grain models have generally been able to reproduce
observed Galactic interstellar polarization curves by small changes
to the size distribution of aligned grains, alignment efficiency,
upper integration limit, etc.  This
is definitely not the case for the M31 curves in that similar changes
to a distribution of astronomical silicates failed to produce
K values of the observed amplitude.  Even more extreme modifications to
the size distribution of silicate particles (use of ``flat''
distributions, very narrow size ranges, etc.) tended to produce K values that
still remained 30-40\% too low, at best.  As a result, it would appear
that there is something intrinsically different (and interesting) about the 
grains
producing the observed polarization in M31 from those in our Galaxy. We
intend to investigate this behavior in much greater depth after our
sample of M31 globulars has been increased.

Previously, only S78 had enough wavelength coverage for a Serkowski fit.
Our new data, which have better wavelenth coverage and better S/N,
show that $\lambda_{Max}$$\sim$4775 \AA\ for S78, very different from 
that found by Martin \& Shawl (1982). 
Their value of 3500 \AA\ is at the extreme low end of $\lambda_{max}$ values measured in 
the 
Galaxy 
(e.g., Serkowski et al. 1975). 
The unusual sightline toward HD 210121, which has one of the 
lowest
R$_V$ values ever measured, has 
$\lambda_{max}\sim$3800 \AA~(Larson, Whittet, \& Hough 1996).
None of the $\sim$20 sightlines measured in the LMC show 
$\lambda_{max}$ values below 4900 \AA~(Clayton et al. 1983). 
Only a few sightlines have been observed in the SMC and the signal-to-noise 
is relatively 
low but $\lambda_{max}$ lies in the normal Galactic range
 (Rodrigues et al. 1997).

The fits for the four sightlines in M31
reveal values of $\lambda_{max}$ ranging from 4800 to 5500 \AA.
These are very average
values of $\lambda_{max}$ as measured in the 
Galaxy and the Magellanic Clouds.
It has been found that R$_V\sim$ 5.5$\lambda_{max}$ 
(Serkowski et al. 1975, Whittet \& van Breda 1978; Clayton \& Mathis 1988).
The range of $\lambda_{max}$ measured in M31 corresponds to R$_V$ values of 2.7 to 3.1.
The range in R$_V$ seen in the Galaxy is 2.5 to 5.5.
So, at least for this very small sample, M31 does not show evidence 
for sightlines with greater than average grain size, if the nature of 
the grains is the same as for the Galaxy. 
This result is in contrast to the paucity of small grains inferred 
from 
IRAS colors (Xu \& Helou 1994).

The polarizing efficiency of the M31 dust grains can be estimated using
the ratio of peak linear polarization to color excess, p$_{max}$/E(B-V).  
While the quantity is often equated with the degree of grain
alignment efficiency, it is also a function of the grain's non-spherical
nature as well as the angle between the alignment axis and the line of
sight.  However, for brevity, we simply refer to the value as the
``efficiency ratio.''
The maximum value of this ratio found in the Galaxy is 9\% mag$^{-1}$ 
(Serkowski et al. 1975; Clayton \& Cardelli 1988). 
We corrected the estimated reddenings for a Galactic foreground of E(B-V)=0.1 mag 
(Barmby et al. 2000).
The values of p$_{max}$/E(B-V) for our sample are given in Table 1. They range from 
7.1 to 15.3\% mag$^{-1}$.
These values generally support the results of Martin \& Shawl (1982) and 
imply a similar polarization efficiency in M31 and the Galaxy.
The values of p$_{max}$/E(B-V) for our sample are all near the maximum value but
some bias is expected since sightlines known to have significant interstellar 
polarization were selected
for inclusion in this sample.
The polarization efficiency toward S233 seems very high, but the amount of
reddening is fairly small and the uncertainty is a
considerable fraction of the E(B-V) value itself. 

Baade 327 is of interest as it is both the most luminous globular 
cluster known and
also the most reddened globular cluster in M31 (Barmby, Perrett, \& Bridges 2002).
van den Bergh (1968) suggested that it is strange that the brightest globular cluster 
in M31 would 
also be the most reddened. This coincidence led to speculation that perhaps the value
of R$_V$ might be small along this sightline resulting in less total
extinction, A$_V$. Our result for Baade 327, $\lambda_{max}\sim$ 5450 \AA\, implies
that the dust along this sightline has the Galactic average, R$_V\sim$3.1.
Even if R$_V$ = 2.5, as suggested by Kron \& Mayall (1960), 
Baade 327 only drops to the second 
most luminous globular cluster
(Barmby et al. 2002). 
They also point out that the most reddened globular clusters lie close 
to the center of M31 and those close to the center of M31 seem to be brighter. 
The polarization efficiency toward Baade 327 is tantalizingly high,
particularly given its large color excess and its
anticipated favorable viewing geometry (i.e., the angle between alignment
and line of sight) (Martin \& Shawl 1979, 1982).  Clearly, additional
observations of M31 globular clusters would provide further insight
into the dust environment in that galaxy.

\acknowledgements
We thank Pauline Barmby for providing us with reddening estimates for the 
globular clusters.
 We also thank Xiaohui Fan and Gary Schmidt for the use
of the CCD Spectropolarimeter at the MMT during periods when weather conditions
were unsuitable for the primary science program.
This study has been supported by NASA ATP grant, NAG5-9203.

\clearpage

\begin{figure*}
\figurenum{1}
\epsscale{1.0}

\caption{Spectropolarimetric observations of M31 globular clusters. 
The upper panels show the foreground corrected 
polarization, the middle panels show position angle and the bottom 
panels show the spectra.
Open squares represent the MMT data,the stars represent 
the WIYN data and the triangles represent the Martin \& Shawl (1982) data. 
The error bars are 1$\sigma$. 
The lines drawn represent the best Serkowski fits to the MMT data only. 
See text.}
\end{figure*}

\begin{figure*}
\figurenum{2}
\epsscale{0.75}
\caption{The Serkowski fit parameters, K and $\lambda_{max}$ are plotted for the 
four M31 globular clusters. The error bars are 1$\sigma$. The line represents
the K = 1.66$\lambda_{max}$ relationship that holds for Galactic interstellar polarization
(Whittet et al. 1992).}
\end{figure*}

\clearpage

\begin{deluxetable}{lllllll}
\tablewidth{0pc} 
\tablenum{1}
\tablecaption{Observed M31 Globular Clusters}
\tablehead{ 
\colhead{Cluster\tablenotemark{a}} & \colhead{$V$\tablenotemark{b}} & \colhead{$E(B-V)$\tablenotemark{b}} &
\colhead{$p_{max}$(\%)\tablenotemark{c}} & \colhead{$\lambda_{max}$ (\AA)\tablenotemark{c}} & \colhead{K\tablenotemark{c}}
& \colhead{p$_{max}$/E(B-V)(\% mag$^{-1}$)\tablenotemark{d}}
}
\startdata 
S78&14.26&0.36$\pm$0.04&1.96$\pm$0.03&4775$\pm$250&1.32$\pm$0.28&7.5$\pm$0.15\\
S150&15.44&0.48$\pm$0.06&3.19$\pm$0.05&5200$\pm$225&1.78$\pm$0.46&8.4$\pm$0.16\\
S233&15.38&0.23$\pm$0.05&1.99$\pm$0.03&5275$\pm$175&2.38$\pm$0.49&15.3$\pm$0.39\\
Baade 327&16.71&1.38$\pm$0.02&9.14$\pm$0.12&5450$\pm$175&2.22$\pm$0.35&7.1$\pm$0.02\\
\enddata
\tablenotetext{a}{Vetesnik (1962); Sargent et al. (1977)}
\tablenotetext{b}{Total reddening as estimated by Barmby et al. (2000)}
\tablenotetext{c}{Based on MMT data only}
\tablenotetext{d}{Galactic foreground reddening of E(B-V)=0.1 subtracted.}
\end{deluxetable}

\begin{deluxetable}{lllll}
\tablewidth{0pc} 
\tablenum{2}
\tablecaption{S78-MMT}
\tablehead{ 
\colhead{$\lambda$-bin (\AA)} & \colhead{p(\%)} & \colhead{$\theta$} &
\colhead{p$_{corr}$} &
\colhead{$\theta_{corr}$} 
}
\startdata 
4125   &   1.78$\pm$    0.36 &  41.5 $\pm$   5.4 &   1.86 &  35.76\\
4375   &   2.00  $\pm$  0.09  &  51.8 $\pm$   1.5  &   1.94  &  46.32\\
4625   &   1.89  $\pm$  0.08  &  50.0 $\pm$   1.1  &   1.86  &  44.09\\
4875   &   1.97 $\pm$   0.05  &  49.3 $\pm$   0.9  &   1.95  &  43.58\\
5125   &   2.04 $\pm$   0.04  &  51.5 $\pm$   0.6  &   1.99  &  45.89\\
5375   &   2.01 $\pm$   0.04  &  51.5 $\pm$   0.6  &   1.96  &  45.81\\
5625   &   1.90 $\pm$   0.04  &  52.8 $\pm$   0.7  &   1.83  &  46.72\\
5875   &   1.96 $\pm$   0.04  &  51.5  $\pm$  0.4  &   1.91  &  45.63\\
6125   &   1.92 $\pm$   0.03  &  51.9  $\pm$  0.5  &   1.87  &  45.98\\
6375   &   1.76 $\pm$   0.03  &  51.8  $\pm$  0.7  &   1.71  &  45.41\\
6625   &   1.75 $\pm$   0.03  &  53.6 $\pm$   0.4  &   1.68  &  47.25\\
6875   &   1.77 $\pm$   0.04  &  53.3 $\pm$   0.5   &  1.70  &  47.12\\
7125   &   1.60 $\pm$   0.04 &  52.2 $\pm$   0.6   &  1.55  &  45.38\\
7375   &   1.50 $\pm$   0.05  &  51.9 $\pm$   0.8   &  1.46  &  44.79\\
7625   &   1.54  $\pm$  0.06  &  50.9 $\pm$   1.0  &   1.51  &  44.14\\
7875   &   1.49 $\pm$   0.05  &  52.2 $\pm$   1.0  &   1.44  &  45.29\\
8125   &   1.45 $\pm$   0.07  &  53.0 $\pm$   1.4  &   1.39  &  46.09\\
\enddata
\end{deluxetable}

\begin{deluxetable}{lllll}
\tablewidth{0pc} 
\tablenum{3}
\tablecaption{S78-WIYN}
\tablehead{ 
\colhead{$\lambda$-bin (\AA)} & \colhead{p(\%)} & \colhead{$\theta$} &
\colhead{p$_{corr}$} &
\colhead{$\theta_{corr}$} 
}
\startdata 
5500 &   2.37  $\pm$ 0.23& 61.7 $\pm$ 2.8 & 2.18 &  57.31\\
  6156 &   1.85 $\pm$ 0.19&  46.7 $\pm$ 2.9 &   1.87  & 40.54\\
  6469  &  1.81 $\pm$ 0.11&  52.1 $\pm$ 1.7 &  1.76&   45.74\\
  6781  &  1.82 $\pm$ 0.09&  52.4 $\pm$ 1.4 &   1.76 &  46.18\\
  7094 &   1.74 $\pm$ 0.08&  52.5 $\pm$ 1.3 &   1.68 &  46.09\\
  7406 &   1.81 $\pm$ 0.07&  53.6 $\pm$ 1.1&   1.74 &  47.58\\
  7719  &  1.60 $\pm$ 0.08&  57.0 $\pm$ 1.4&  1.49 &  50.34\\
  8031 &   1.42 $\pm$ 0.08&  61.4 $\pm$ 1.6&  1.26  & 54.30\\
  8344  &  1.44 $\pm$ 0.08&  61.0 $\pm$  1.6&   1.29 &  54.16\\
\enddata
\end{deluxetable}

\begin{deluxetable}{lllll}
\tablewidth{0pc} 
\tablenum{4}
\tablecaption{S150}
\tablehead{ 
\colhead{$\lambda$-bin (\AA)} & \colhead{p(\%)} & \colhead{$\theta$} &
\colhead{p$_{corr}$} &
\colhead{$\theta_{corr}$} 
}
\startdata
4125  &    2.70 $\pm$   0.44 &  36.7 $\pm$   5.1  &  2.82 &  33.12\\
4375  &    2.60 $\pm$   0.16 &  34.3 $\pm$   1.6  &  2.76 &  30.60\\
4625  &    3.11 $\pm$   0.09 &  36.3 $\pm$   1.0 &  3.25 &  33.00\\
4875  &    2.75 $\pm$   0.09 &  36.4  $\pm$  0.7 &  2.89 &  32.68\\
5125  &    3.12 $\pm$   0.05 &  34.9  $\pm$  0.6  &  3.28 &  31.63\\
5375  &    2.90 $\pm$   0.07 &  35.4 $\pm$   0.6  &  3.05 &  31.81\\
5625  &    2.97 $\pm$   0.06 &  35.0 $\pm$   0.4  &  3.13 &  31.51\\
5875  &    3.21 $\pm$   0.05 &  32.7 $\pm$   0.5  &  3.39 &  29.63\\
6125  &    2.81 $\pm$   0.05 &  35.5 $\pm$   0.4  &  2.96 &  31.84\\
6375  &    2.69 $\pm$   0.05 &  35.4 $\pm$   0.4  &  2.84 &  31.64\\
6625  &    2.62 $\pm$   0.04 &  34.7 $\pm$   0.5  &  2.78 &  30.92\\
6875  &    2.66 $\pm$   0.04 &  34.5 $\pm$   0.6  &  2.82 &  30.87\\
7125  &    2.51 $\pm$   0.05 &  35.3 $\pm$   0.5  &  2.66 &  31.43\\
7375  &    2.35 $\pm$   0.06&   35.8 $\pm$   0.8  &  2.49 &  31.76\\
7625  &    2.34 $\pm$   0.06 &  34.8  $\pm$  0.9  &  2.49 &  30.88\\
7875  &    2.27 $\pm$   0.08 &  36.8  $\pm$  1.2  &  2.39 &  32.75\\
8125  &    2.18 $\pm$   0.13 &  33.1  $\pm$  1.4  &  2.34 &  29.18\\
\enddata
\end{deluxetable}

\begin{deluxetable}{lllll}
\tablewidth{0pc} 
\tablenum{5}
\tablecaption{S233}
\tablehead{ 
\colhead{$\lambda$-bin (\AA)} & \colhead{p(\%)} & \colhead{$\theta$} &
\colhead{p$_{corr}$} &
\colhead{$\theta_{corr}$} 
}
\startdata
4125  &    1.82  $\pm$  0.42 &  53.6 $\pm$   5.7  &  1.74 &  47.76\\
4375  &    2.03 $\pm$   0.18 &  44.9 $\pm$   2.5  &  2.06 &  39.59\\
4625  &    1.84 $\pm$   0.12 &  44.8 $\pm$   1.8  &  1.88 &  38.82\\
4875   &   1.90 $\pm$   0.11 &  43.5 $\pm$   1.6 &   1.96 &  37.67\\
5125  &    1.92 $\pm$   0.08 &  41.3 $\pm$   1.2  &  2.01 &  35.64\\
5375  &    1.94 $\pm$   0.07 &  44.0 $\pm$   1.4  &  1.99 &  38.20\\
5625  &    1.74 $\pm$   0.10 &  42.6 $\pm$   1.1  &  1.82 &  36.21\\
5875  &    1.78 $\pm$   0.07 &  44.6 $\pm$   1.1  &  1.83 &  38.25\\
6125  &    1.85 $\pm$   0.05 &  43.4 $\pm$   0.9  &  1.91&   37.48\\
6375  &    1.84 $\pm$   0.06 &  43.0 $\pm$   1.0  &  1.91 &  37.11\\
6625  &    1.74 $\pm$   0.06  & 44.5 $\pm$   1.0  &  1.79 &  38.25\\
6875  &    1.63 $\pm$   0.07 &  47.2 $\pm$   1.2 &   1.64 &  40.57\\
7125  &    1.62 $\pm$   0.06&  44.0 $\pm$   1.3 &   1.67&   37.53\\
7375   &   1.52 $\pm$   0.09 &  45.5 $\pm$   1.4 &   1.56 &  38.63\\
7625   &   1.36 $\pm$   0.08 &  52.3 $\pm$   1.8  &  1.31 &  44.58\\
7875  &    1.34 $\pm$   0.14 &  37.6 $\pm$   2.4  &  1.47 &  30.82\\
8125   &   1.02 $\pm$   0.15 &  51.6  $\pm$   3.8  &  1.00 &  41.77\\
\enddata
\end{deluxetable}

\begin{deluxetable}{lllll}
\tablewidth{0pc} 
\tablenum{6}
\tablecaption{Baade 327}
\tablehead{ 
\colhead{$\lambda$-bin (\AA)} & \colhead{p(\%)} & \colhead{$\theta$} &
\colhead{p$_{corr}$} &
\colhead{$\theta_{corr}$} 
}
\startdata
4125  &    5.72 $\pm$   3.26 &  74.0 $\pm$   9.7 &   5.40 &  73.01\\
4375  &    6.49 $\pm$   1.68 &  45.8  $\pm$  5.2  &  6.50 &  44.10\\
4625  &    7.31 $\pm$   1.05 &  54.5 $\pm$  3.3  &  7.19 &  53.02\\
4875  &    8.81 $\pm$   0.74 &  58.3 $\pm$   1.9 &   8.64 &  57.10\\
5125  &    9.09  $\pm$  0.45 &  53.0 $\pm$   1.3 &   8.99 &  51.78\\
5375  &    9.37 $\pm$   0.31 &  56.1 $\pm$   0.8 &   9.21 &  55.86\\
5625  &    9.57 $\pm$   0.23 &  53.9 $\pm$   0.7 &   9.46 &  52.73\\
5875  &    9.18 $\pm$   0.20 &  53.1 $\pm$   0.6 &   9.08 &  51.85\\
6125  &    9.04 $\pm$   0.18 &  53.8 $\pm$   0.4 &   8.93 &  52.58\\
6375  &    8.63 $\pm$   0.14 &  54.7 $\pm$   0.5 &   8.51 &  53.42\\
6625  &    8.62 $\pm$   0.15 &  54.9 $\pm$   0.4 &   8.50 &  53.63\\
6875  &    8.37 $\pm$   0.13 &  55.2 $\pm$   0.4 &   8.25 &  54.00\\
7125  &    7.74 $\pm$   0.13 &  54.9 $\pm$   0.5 &   7.62 &  53.59\\
7375  &    7.33 $\pm$   0.13 &  53.4 $\pm$   0.5 &   7.23 &  52.04\\
7625  &    7.28 $\pm$   0.12 &  54.4 $\pm$   0.5 &   7.17 &  53.05\\
7875  &    7.15 $\pm$   0.19 &  54.0 $\pm$   0.5 &   7.05 &  52.61\\
8125  &    6.72 $\pm$   0.22 &  55.1 $\pm$   0.8 &   6.60 &  53.66\\
\enddata
\end{deluxetable}


\clearpage

\begin{references}
\reference{} Barmby, P., Huchra, J.P., Brodie, J.p., Forbes, D.A., Schroder, L.L., 
\& Grillmair, C.J. 2000, AJ, 119, 727
\reference{} Barmby, P., Perrett, K.M., \& Bridges, T.J. 2002, MNRAS, 329, 461
\reference{} Bianchi, L., Clayton, G.C., Hutchings, J.B., Massey, P., \& Bohlin, R.C. 1996, 
ApJ, 
471, 203
\reference{} Cardelli, J. A., Clayton, G. C., \& Mathis 1989, ApJ, 345, 245 (CCM)
\reference{} Clayton, G.C., \& Cardelli, J.A. 1988, AJ, 96, 695
\reference{} Clayton, G.C., Green, J., Wolff, M.J., Zellner, N.E.B., Code, A.D., \& 
Davidsen, A.F. 
1996, Ap.J., 460, 313
\reference{} Clayton, G.C., Martin, P.G., \& Thompson, I. 1983, ApJ, 265, 194
\reference{} Clayton, G.C., \& Mathis, J.S. 1988, ApJ, 327, 911
\reference{} Clayton, G.C., Wolff, M.J., Allen, R.G., \& Lupie, O.L. 1995, ApJ, 445, 947
\reference{} Freedman, W.L., \& Madore, B.F. 1990, ApJ, 365, 186 
\reference{} Gordon, K. D. \& Clayton, G.C. 1998, ApJ, 500, 816
\reference{} Gordon, K.D., Clayton, G.C., Misselt, K. A., Landolt, A.U., \& Wolff, M.J. 
2003, ApJ, 594, 279
\reference{} Hiltner, W.A. 1958, ApJ, 128, 9
\reference{} Iye, M., \& Richter, O.-G. 1985, A\&A, 144, 471
\reference{} Kron, G.E., \& Mayall, N.U. 1960, AJ, 65, 581
\reference{} Larson, K.A., Whittet, D.C.B., \& Hough, J.H. 1996, ApJ, 472, 755
\reference{} Lequeux, J. 2000, The Interstellar Medium in M31 and M33 Proceedings 232, 
Eds. E. M. Berkhuijsen, R. Beck,\& R. A. M. Walterbos, (Shaker:Aachen), p. 63-68
\reference{} Martin, P. G., Clayton, G.C., \& Wolff, M.J. 1999, ApJ, 510, 905
\reference{} Martin, P.G. \& Shawl, S.J. 1979, ApJ, 231, L57
\reference{} Martin, P.G. \& Shawl, S.J. 1982, ApJ, 253, 86
\reference{} Massey, P., Armandroff, T. E., Pyke, R., Patel, K., \& Wilson, C. D. 
1995, AJ, 110, 2715
\reference{} Massey, P., Strobel, K., Barnes, J. V., \& Anderson, E. 1988, ApJ, 328, 315
\reference{} Misselt, K., Clayton, G.C.,\& Gordon, K.D. 1999, ApJ, 515, 128
\reference{} Nedialkov, P., \& Veltchev, T. 1999, astro-ph/9911262
\reference{} Nordsieck, K. H., \& Harris, W. 1996, in ASP Conf. Ser. 97, Polarimetry 
of the Interstellar Medium, ed. W. G. Roberge \& D. C. B. Whittet (San Francisco: ASP), 100
\reference{} Rodrigues, C.V., Magalhaes, A.M., Coyne, C.V., S.J., \& Piirola, V. 1997, ApJ, 
485, 618
\reference{} Sargent, W.L.W., Kowal, C.T., Hartwick, F.D.A., \& van den Bergh, S. 1977, AJ, 
82, 947
\reference{} Schmidt, G.D., Elston, R., \& Lupie, O.L. 1992, AJ, 104, 1563
\reference{} Schmidt, G. D., Stockman, H. S., \& Smith, P. S. 1992, ApJ, 398, L57
\reference{} Searle, L. 1983, Carnegie Inst. of Washington Yearbook, 82, 622
\reference{} Serkowski, K., Mathewson, D. S., \& Ford, V. L. 1975, ApJ, 196, 261
\reference{} Sharov, A.S., \& Lyutyj, V.M. 1989, Astr. Zh., 66, 462
\reference{} van den Bergh, S. 1968, The Observatory, 88, 168
\reference{} Vetesnik,M. 1962, Bull. Astron. Inst. Czech., 13, 180
\reference{} Whittet, D. C. B., Martin, P. G., Hough, J. H., Rouse, M. F., Bailey, J. A., 
\& Axon, D. J. 1992, ApJ, 386, 562
\reference{} Whittet, D. C. B., \& van Breda, I.G. 1978, A\&A, 66, 57
\reference{} Wilking, B. A., Lebofsky, M. J., Martin, P. G., Rieke, G. H., \& Kemp, J. C. 1980, 
ApJ, 235, 905
\reference{} Wilking, B. A., Lebofsky, M. J., \& Rieke, G. H. 1982, AJ, 87, 695
\reference{} Wolff, M.J., Clayton, G.C., Martin, P.G., \& Schulte-Ladbeck, 
R.E. 1994, Ap.J., 423, 412
\reference{} Wolff, M.J., Nordsieck, K.H., \& Nook, M.A. 1996, AJ, 111, 856
\reference{} Xu, C, \& Helou, G. 1994, ApJ, 426, 109



\end{references}
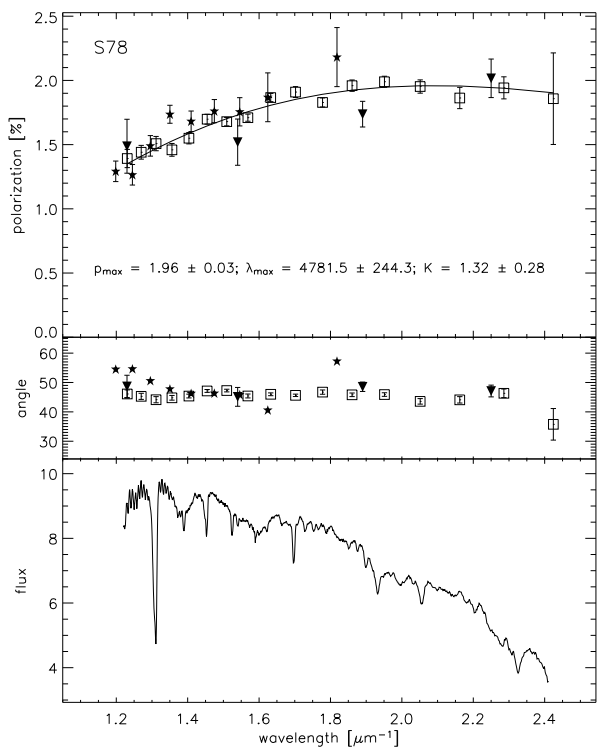
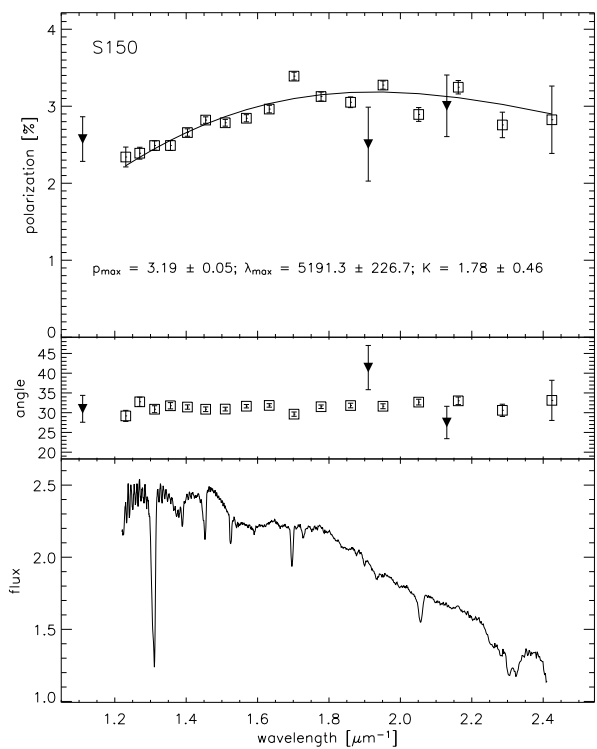
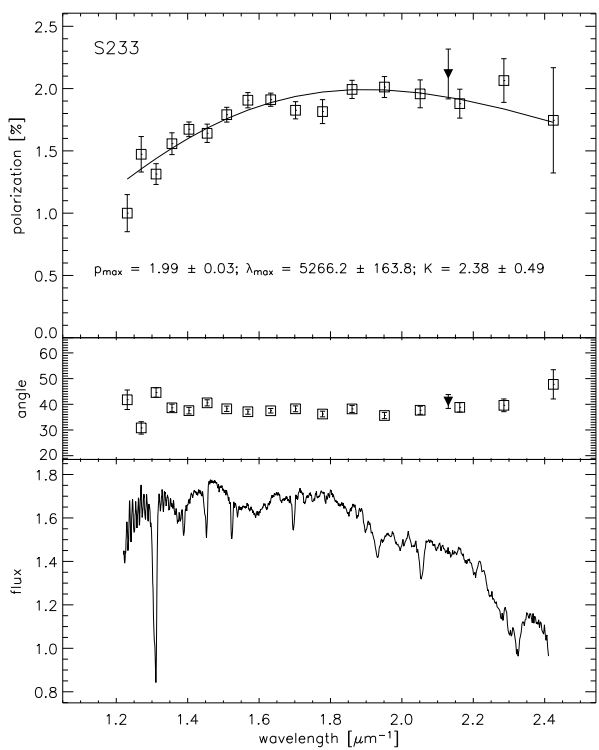
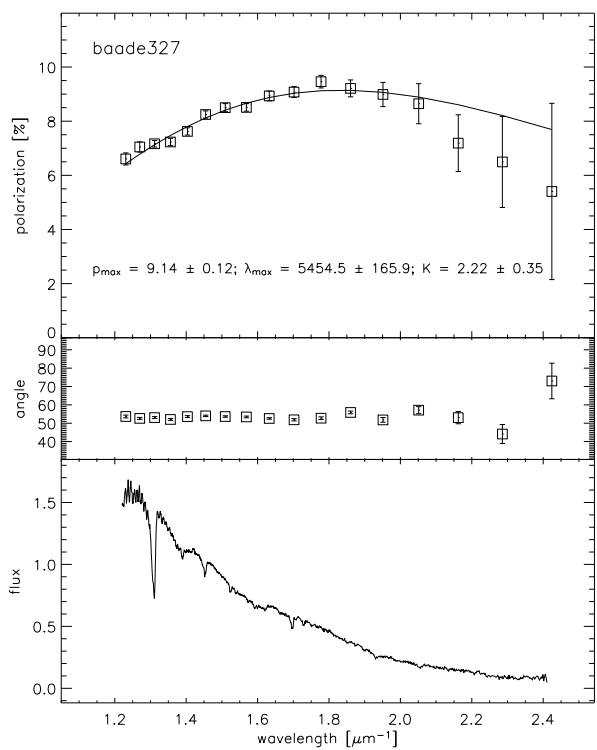
\end{document}